\documentclass[11pt]{article}
\usepackage{cite}
\usepackage{mdframed}
\usepackage{epsfig,subfigure}
\usepackage{amssymb}
\usepackage[fleqn]{amsmath}
\usepackage{color}
\usepackage{arydshln}
\def\QED{\mbox{\rule[0pt]{1.5ex}{1.5ex}}}

\usepackage{graphicx}
\usepackage{color}
\usepackage[dvipsnames]{xcolor}

\definecolor{armygreen}{rgb}{0.29, 0.33, 0.13}

\usepackage{amssymb}
\usepackage{amsmath,amsfonts,amssymb}
\usepackage{verbatim}
\usepackage{stfloats}
\usepackage[bookmarks=false]{}
\usepackage{algorithm}
\usepackage{algcompatible}

\newtheorem{theorem}{Theorem}

\newtheorem{definition}{Definition}
\newtheorem{lemma}{Lemma}

\newtheorem{remark}{Remark}

\newtheorem{example}{Example}
\newcommand\blfootnote[1]{%
  \begingroup
  \renewcommand\thefootnote{}\footnote{#1}%
  \addtocounter{footnote}{-1}%
  \endgroup
}

\begin{document}
\date{}
%\title{The Capacity of Anonymous Communications
%%Anonymous Communications over a Multiple Access Channel
%%\thanks{This work is supported by NSF grants CCF-1317351 and CCF-0963925.}
%}
%\author{\IEEEauthorblockN{Hua Sun}
%\IEEEauthorblockA{Department of Electrical Engineering\\
%University of North Texas, Denton, TX 76203\\
%hua.sun@unt.edu}
%%\and
%%\IEEEauthorblockN{Syed A. Jafar}
%%\IEEEauthorblockA{Center for Pervasive Communications and Computing (CPCC)\\
%%University of California Irvine, Irvine, CA 92697\\
%%syed@uci.edu}
%}

\title{%Blind Interference Alignment for \\ Private Information Retrieval
Conditional Disclosure of Secrets: \\
A Noise and Signal Alignment Approach
%\thanks{This work is supported by NSF grants CCF-1317351 and CCF-0963925.}
}
\author{\normalsize Zhou Li and Hua Sun \\
}

\maketitle

\blfootnote{%This paper will be presented in part at ISIT 2018. 
Zhou Li (email: zhouli@my.unt.edu) and Hua Sun (email: hua.sun@unt.edu) are with the Department of Electrical Engineering at the University of North Texas. }

\maketitle

\begin{abstract}
In the conditional disclosure of secrets (CDS) problem, Alice and Bob (each holds an input and a common secret) wish to disclose, as efficiently as possible, the secret to Carol if and only if their inputs satisfy some function.
%The (symmetric) capacity of CDS is the maximum number of bits of the secret that can be securely disclosed per bit of total communication, when the number of bits communicated from Alice and Bob to Carol is the same for all input values. We characterize the necessary and sufficient condition for the extreme case where the capacity of CDS is the highest and is equal to $1/2$. For the simplest instance where the capacity is smaller than $1/2$, we show that the linear capacity is $2/5$.
The capacity of CDS is the maximum number of bits of the secret that can be securely disclosed per bit of total communication. We characterize the necessary and sufficient condition for the extreme case where the capacity of CDS is the highest and is equal to $1/2$. For the simplest instance where the capacity is smaller than $1/2$, we show that the linear capacity is $2/5$.

%symmetric
%extreme
\end{abstract}

\newpage

\allowdisplaybreaks
\section{Introduction}
In a seminal work \cite{shannon1949}, Shannon introduced the notion of perfect security (also known as information theoretic security) based on statistical independence and established the fundamental limits %on communication and randomness cost 
of a single-user secure communication system. While \cite{shannon1949} provided an elegant theoretical foundation for cryptography, the optimal solutions are deemed too inefficient to implement in practice \cite{Crypto_book}. Cryptographers therefore relax the stringent requirement of perfect security to computational security, defined based on indistinguishability with limited computation power. Most existing commercial security protocols are built on computational security.

Modern secure communication systems naturally involve multiple users. Interestingly, for multi-user secure communication systems, solutions based on perfect security are not necessarily less efficient when compared to those based on computational security (e.g., see \cite{Devet_Goldberg} for a specific context of private information retrieval). As such, there is much potential for perfect security in multi-user networks, especially considering the fact that multi-user security protocols based on both perfect and computational security criteria are primarily studied in academia and large-scale practical implementations are still rare. It is thus imperative to understand the fundamental limits of perfect security in multi-user networks, which has been studied in the cryptography and theoretical computer science communities \cite{cramer_damgård_nielsen_2015}, although typically not using information theoretic tools. Due to the increasing importance of security in modern communication systems, it has also recently become one of the focuses for the information theory community \cite{liang2009information, bloch2011physical, yener2015wireless}, where both classical cryptography formulations are studied \cite{Sun_Jafar_PIR, Banawan_Ulukus, Lee_Abbe, Data_Prabhakaran_Prabhakaran, Zhou_Sun_Fu, Zhao_Sun_SMP} and new models are introduced \cite{yu2018lagrange, chang2018capacity, Sun_Anonymous, Tahmasebi_Maddah, Wang_Banawan_Ulukus}. The goal of this paper is to use information theoretic tools to study a canonical theoretical computer science problem (i.e., a cryptographic primitive) - conditional disclosure of secrets (CDS) \cite{SymPIR, Gay_Kerenidis_Wee, Applebaum_Arkis_Raykov_Vasudevan}.

\begin{figure}[h]
\begin{center}
\includegraphics[width= 4 in]{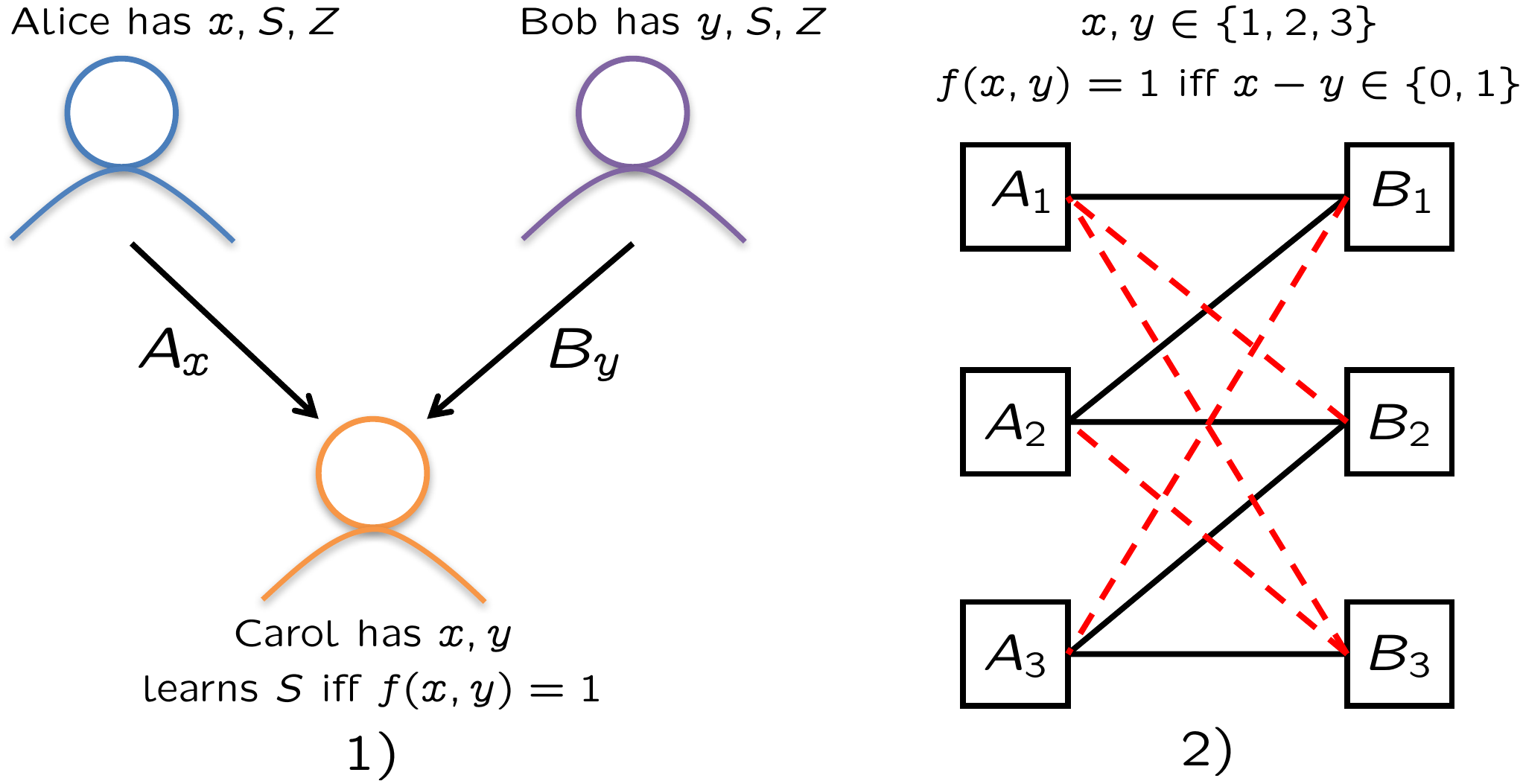}
\vspace{-0.15in}
\caption{\small 1). The CDS problem. 2) An example of $f(x,y)$ represented by a bipartite graph. Nodes in the left (right) column are the signals from Alice (Bob) for various input values. From pair of nodes connected by a solid black edge (i.e., $f(x,y) = 1$), Carol can decode $S$; from pair of nodes connected by a dashed red edge (i.e., $f(x,y) = 0$), Carol learns nothing about $S$.
%$x, y$ are inputs, $S$ is the secret, $Z$ is a noise variable and $A_x, B_y$ are output signals.
}
\label{fig:prob}
\end{center}
\end{figure}

In the CDS problem (see Fig.~\ref{fig:prob}), Alice and Bob hold inputs $x$ and $y$ respectively, in addition to a common secret $S$. Alice and Bob wish to disclose the secret $S$ to Carol if their inputs $x, y$ satisfy some function $f$, i.e., when $f(x, y) = 1$. Otherwise $f(x, y) = 0$, absolute no information is revealed to Carol in the information theoretic sense. A common noise variable $Z$ is available to Alice and Bob to assist the task, while Carol is fully ignorant of $Z$. Alice and Bob send signals $A_x$ and $B_y$ respectively to Carol. The aim is to find an efficient communication protocol, i.e., we wish to minimize the number of bits contained in $A_x$ and $B_y$. 

\vspace{-0.1in}

\subsection{Motivation}
The CDS problem is a minimal model that captures the challenges of characterizing the communication cost of security in multi-user networks. Note that if there is no security constraint, the problem is trivial as either Alice or Bob may directly send the secret to Carol. However, once the security constraint is included, the optimal communication cost of the CDS problem immediately becomes one of the notable open problems in information theoretic cryptography \cite{Vaikuntanathan}. Further, considering that there are only three users in the CDS problem, we find it to be a simplest intriguing theoretical model and are interested in understanding its fundamental communication limits.

Beyond the theoretical value, the CDS problem is also relevant in modeling several interesting secure communication scenarios. One may interpret $x$ and $y$ as the queries sent from the user Carol to two distributed non-communicating servers, Alice and Bob, respectively. The signals $A_x$ and $B_y$ are the answers from the servers that enable the user to obtain the desired data, $S$. The security condition of $f(x, y) = 1$ is to ensure that data retrieval is successful if the queries are qualified while for unqualified queries, nothing is revealed. Note that the distributed servers are non-communicating so that Alice only knows $x$ and Bob only knows $y$. We then need a mechanism for Alice and Bob to produce answers without knowing the other query. In fact, the CDS problem was introduced first in the context of symmetric private information retrieval \cite{SymPIR}, exactly motivated by this need of providing distributed data access service with protection under unqualified (malicious) queries. Another interesting application may be seen as follows. Alice and Bob wish to share the secret (e.g., a business plan) with Carol if and only if they wish to collaborate, and $f(x,y)$ captures the condition under which they {\em agree} to collaborate.

From a different perspective, the CDS problem could be viewed as a secure data storage system over a bipartite graph (see Fig.~\ref{fig:prob}.2). The nodes in the graph are the storage variables and there are two types of edges, where from the pair of nodes connected by one type of edge, the secret is recoverable and otherwise, from the other type of edge, no information is disclosed. As such, the CDS problem is meant to provide fine-grained access control for encrypted data, where the access structure\footnote{It is interesting to compare CDS with a related problem secret sharing \cite{Beimel_Survey}, where a secret is distributed over multiple nodes in a way that if and only if a set of nodes belong to certain pre-defined sets, the secret can be reconstructed. The main difference is that for secret sharing, either perfect reconstruction or zero leakage is guaranteed for {\em any} set of nodes. In contrast, the access structure of CDS is much sparser. For example, for nodes on the same side of the bipartite graph (e.g., $A_1, A_2, A_3$ in Fig.~{\ref{fig:prob}}.2), no condition is placed, i.e., they may or may not be sufficient to recover the secret or part of the secret. Note that this sparsity is an important distinction, e.g., we will develop an alignment view to CDS in this paper, while an alignment view to secret sharing is not available yet. In addition, qualified sets in CDS only have 2 nodes while any number of nodes is allowed in secret sharing.} may be very diverse depending on the underlying graph (i.e., $f(x,y)$). For other applications of CDS, we refer to \cite{Gay_Kerenidis_Wee, Laur_Lipmaa, Liu_Vaikuntanathan_Wee, Applebaum_Arkis_Raykov_Vasudevan} and references therein.

\subsection{Comparison to Previous Approach}
In cryptography and theoretical computer science communities, the typical formulation of the CDS problem is as follows \cite{SymPIR, Gay_Kerenidis_Wee, Applebaum_Arkis_Raykov_Vasudevan, Vaikuntanathan, Laur_Lipmaa, Liu_Vaikuntanathan_Wee, Applebaum_Vasudevan}.
\begin{itemize}
\item The secret $S$ has {\em 1 bit}. The communication cost (i.e., the number of bits in $A_x, B_y$) is measured as order functions of the input size (the logarithm of the number of possibilities of inputs $x, y$). So the studied question is - how does the communication cost of disclosing a one-bit secret {\em scale with the complexity of the function $f(x,y)$}?
\item Implicit to the above formulation is that the proposed protocols must work for {\em all} functions $f(x,y)$. In other words, the considered setting is the {\em worst} case scenario that targets at the most challenging $f(x,y)$.
\end{itemize}

In contrast, in this work we will take a Shannon theoretic formulation.
\begin{itemize}
\item We allow the {\em secret size to scale to infinity while the function $f(x,y)$ is fixed}. Our metric is the communication rate, which is defined as the ratio of the secret size to the number of bits communicated to Carol. So our question is - what is the maximum number of bits that can be secretly disclosed, per bit of total communication?
\item Regarding the function $f(x,y)$, we are interested in the {\em instance optimal} setting, i.e., for a fixed instance of $f(x,y)$, what is the optimal communication strategy?
\end{itemize}

\subsection{Main Contribution and Technique}
In this work, we mainly consider the best cases of $f(x,y)$, i.e., when the communication rate is the highest. As long as the security constraint is not empty for any input value (i.e., for any $x$ ($y$), there exist some $y$ ($x$) such that $f(x,y) = 0$), the size of $A_x, B_y$ cannot be smaller than the secret size (as each of $A_x$ and $B_y$ must be independent of the secret by itself). For all such non-degenerate cases, the rate cannot be larger than $1/2$, because to disclose 1 bit of the secret, both Alice and Bob must communicate $1$ bit to Carol (then the total communication must be at least $2$ bits). Our first main result is a complete characterization of all instances of $f(x,y)$ such that the capacity of CDS is $1/2$ (see Theorem \ref{thm:1/2rate}). The characterization is stated in terms of the graph theoretic properties of $f(x,y)$. Our second main result is the linear capacity characterization of the simplest CDS instance such that its capacity is smaller than $1/2$ (see Theorem \ref{thm:2/5rate}). Interestingly, once we go beyond the best case of capacity $1/2$, the problem becomes significantly more challenging and we are only able to settle the linear capacity.

The main results are obtained using an alignment view of the CDS problem, which can be viewed as generalizations and adaptations of interference alignment \cite{Jafar_FnT}. Interference alignment originated in wireless networks \cite{Cadambe_Jafar_int, MMK} and has been applied much beyond the wireless context, e.g., to distributed storage repair \cite{Wu_Dimakis, Shah_IA, Cadambe_Jafar_Maleki_Ramchandran_Suh}, to network coding \cite{meng2014precoding, Han_Wang_Shroff} and index coding \cite{Maleki_Cadambe_Jafar, Sun_Jafar_nonshannon}, and to private information retrieval \cite{Sun_Jafar_BIAPIR, Jia_Sun_Jafar}. Interference alignment aims to let the {\em multiple} undesired signal spaces overlap as much as possible, so as to maximize the number of dimensions left for the desired signal. It is then obvious that in the CDS problem, we only have two objects -  the secret $S$ and the noise $Z$, so there is no interference to say, not to mention multiple interferences. What we develop in this work is a new look of the CDS problem from the perspective of the overlap of the noise spaces and the signal spaces, i.e., noise alignment and signal alignment. Both the converse results and achievable schemes are based on such an alignment argument.  

%\begin{itemize}
%\item {Noise Alignment:} If $f(x,y) = 1$, i.e., from $A_x$ and $B_y$, we may recover the secret, then the noise space in $A_x$ and $B_y$ must overlap with a size that is not smaller than the secret size. Then we need to consider the alignment of noise spaces over non-adjacent signals, i.e., $A_x$ and $B_y$ that are not directly connected.
%
%\item {Signal Alignment:} Given that the noise spaces of $A_x$ and $B_y$ must align to a certain extent, if $f(x,y) = 0$ (i.e., $A_x$ and $B_y$, nothing about the secret is revealed), then 
%\end{itemize}

%\bigskip
%{\it Notation: %$\mathbb{N}$ is the set of natural numbers. 
%For  integers $Z_1, Z_2, Z_1 \leq Z_2$, we use the compact notation $[Z_1:Z_2]=\{Z_1, Z_1+ 1,\cdots, Z_2\}$. %Similarly, $A_{[Z_1:Z_2]} \define \{A_{Z_1},A_{Z_1 + 1},\cdots,A_{Z_2}\}$ for any variable $A$.  
%The notation $X \sim Y$ is used to indicate that random variables $X$ and $Y$ are identically distributed. 
%%The notation $|A|$ is used to denote the cardinality of a set $A$. %when $A$ is a set, and the length of a tuple when $A$ is a tuple.  
%%For sets $S_1, S_2$, we define $S_1/S_2$ as the set  of elements that are in $S_1$ and not in $S_2$. For $n\in\{1,2\}$ we define $\bar{n}$ as the complement of $n$, i.e., $\bar{n}=1$ if $n=2$ and $\bar{n}=2$ if $n=1$.
%}

\section{Problem Statement}\label{sec:model}
Consider a pair of inputs $(x,y)$ from some set $\mathcal{I} \subset \{1,2,\cdots, X\} \times \{1,2,\cdots, Y\}$. Input $x$ is available to Alice and input $y$ is available to Bob. Alice and Bob also both hold a secret $S$ that is comprised of $L$ i.i.d. uniform symbols from a finite field $\mathbb{F}_p$ and an independent common noise variable $Z$ that is comprised of $L_Z$ i.i.d. uniform symbols from $\mathbb{F}_p$. In $p$-ary units,
\begin{eqnarray}
H(S) = L, ~H(Z) = L_Z, ~H(S, Z) = H(S) + H(Z) = L + L_Z. \label{sz_ind}
\end{eqnarray}

Alice and Bob wish to communicate the secret $S$ to Carol if $f(x, y) = 1$, for a globally known binary function $f$, defined over {domain} $\mathcal{I}$. When $f(x,y) = 0$, zero information about $S$ should be revealed. To this end, Alice sends signal $A_x$ and Bob sends signal $B_y$ to Carol. $A_x$ has $L_{A_x}$ symbols from $\mathbb{F}_p$ and $B_y$ has $L_{B_y}$ symbols from $\mathbb{F}_p$. $A_x$ and $B_y$ are functions of $S, Z$, for all $(x,y) \in \mathcal{I}$. %Note that we view the lower-case letter inputs $x, y$ as constants.
\begin{eqnarray}
H(A_x, B_y | S, Z) = 0. \label{det}
\end{eqnarray}

From $A_x, B_y$, Carol can recover $S$ with no error\footnote{The results of this work also hold under the $\epsilon$-error framework.} if $f(x,y) = 1$, and otherwise $f(x,y) = 0$, $A_x, B_y$ must be independent of $S$. For all $(x,y) \in \mathcal{I}$, we have
\begin{eqnarray}
&& (\mbox{Correctness}) ~~H(S | A_x, B_y) = 0,  ~~~~~~~~~\mbox{if}~f(x,y) = 1; \label{dec} \\
&& (\mbox{Security}) ~~~~~~H(S | A_x, B_y) = H(S), ~~~\mbox{otherwise}~f(x,y) = 0. \label{sec}
\end{eqnarray}
The collection of the mappings from $x,y,S,Z$ to $A_x, B_y$ as specified above is called a CDS scheme.

A signal rate tuple $(\frac{L}{L_{A_1}}, \frac{L}{L_{A_2}}, \cdots, \frac{L}{L_{A_X}}, \frac{L}{L_{B_1}}, \cdots, \frac{L}{L_{B_Y}})$ is said to be achievable if there exists a CDS scheme, for which the correctness and security constraints (\ref{dec}), (\ref{sec}) are satisfied. The closure of the set of all achievable signal rate tuples is called the capacity region $\mathcal{C}$. The achievable communication rate characterizes how many symbols of the secret are securely disclosed per symbol of total communication and is defined with respect to the symmetric signal rate tuple
%\footnote{Note that the maximum symmetric rate tuple is equivalent to the max max rate tuple, i.e., $\max_{\mathcal{C}} (\frac{L}{N}, \cdots, \frac{L}{N}) \in \mathcal{C}$ if and only if $\max_{\mathcal{C}} (\frac{L}{L_{A_1}}, \frac{L}{L_{A_2}}, \cdots, \frac{L}{L_{A_X}}, \frac{L}{L_{B_1}}, \cdots, \frac{L}{L_{B_Y}})$ where $\max(A_1,\cdots, A_X, B_1, \cdots, B_Y) = N$.} 
as follows. 
\begin{eqnarray}
R = \frac{L}{2N} ~~\mbox{s.t.}~~(\frac{L}{N}, \cdots, \frac{L}{N}) \in \mathcal{C}.
\end{eqnarray}
The supremum of achievable communicate rates is called the capacity of CDS, $C$. 

The randomness rate specifies how many secret symbols are disclosed per noise symbol and is defined as $R_Z = \frac{L}{L_Z}$. In this work, we focus mainly on the metric of capacity $C$ and allow as much noise as needed, i.e., the randomness rate is unconstrained.

\subsection{Graph Representation of $f(x,y)$}
The function $f(x,y)$ can be equivalently specified by its characteristic undirected bipartite graph $G_f(V, E)$, defined as follows. The vertex set of $G_f$ is comprised of all signals sent from Alice and Bob, i.e., $V = \{A_1, \cdots, A_X, B_1, \cdots, B_Y\}$. As the vertices and the signals have an invertbile mapping, we use vertex and signal {interchangeably} in this paper.
The edge set of $G_f$ is comprised of the unordered pairs $\{A_x, B_y\}$ from the vertex set such that $(x, y) \in \mathcal{I}$. The edges have two types, $t: E \rightarrow \{0,1\}$. For the first type, $\{A_x, B_y\}$ is a solid black edge and is referred to as a {\em qualified edge} if $f(x,y) = 1$ and equivalently $t(A_x, B_y) = 1$; for the second type, $\{A_x, B_y\}$ is a dashed red edge and is referred to as a {\em unqualified edge} if $f(x,y) = 0$ and equivalently $t(A_x, B_y) = 0$ (see Fig.~\ref{fig:prob}.2 for an example).

The following notions of the characteristic graph $G_f$ will be used to state our results. We follow standard graph theory terminologies (e.g., see \cite{Schrijver}). 
\begin{definition}[Qualified/Unqualified Path]
A sequence of distinct connecting qualified (unqualified) edges is called a {\em qualified (unqualified) path}.
\end{definition}

For example, in Fig.~\ref{fig:prob}.2, $P = (\{A_1, B_1\}, \{B_1, A_2\}, \{A_2, B_2\}, \{B_2, A_3\}, \{A_3, B_3\})$ is a qualified path while $P = (\{B_2, A_1\}, \{A_1, B_3\}, \{B_3, A_2\})$ is an unqualified path. Note that a path can be equivalently specified by a sequence of vertices or edges. %For example, the unqualified path $(\{B_2, A_1\}, \{A_1, B_3\}, \{B_3, A_2\})$ can be described as $(B_2, A_1, B_3, A_2)$.

\begin{definition}[Internal Qualified Edge]
A qualified edge that connects two vertices in an unqualified path is called an {\em internal qualified edge}.
\end{definition}

For example, consider the unqualified path $P = (\{B_2, A_1\}, \{A_1, B_3\}, \{B_3, A_2\})$ in Fig.~\ref{fig:prob}.2, which can be equivalently specified by a vertex sequence $(B_2, A_1, B_3, A_2)$. The qualified edge $\{B_2, A_2\}$ is an internal qualified edge.

\begin{definition}[Qualified Component]
A {\em qualified (connected) component} is a maximal induced subgraph of $G_f$ such that any two vertices in the subgraph are connected by a qualified path.
\end{definition}

In this work, to avoid degenerate settings and to simplify the presentation of results\footnote{Note that a degenerate setting can be converted to a non-degenerate one. Consider any vertex $v$ that is {connected to} only qualified edges. In other words, this vertex has no security constraint. Then we may set the signal $v$ to be the secret $S$ and eliminate $v$. Repeating the same procedure for all such vertices, we have a non-degenerate setting.}, we restrict ourselves to functions $f(x,y)$ such that the security constraint (\ref{sec}) is not empty for any individual $x$ and any individual $y$.

\begin{definition}[Non-degenerate Condition]
A CDS instance, described by the characteristic graph $G_f(V,E)$ is called {\em non-degenerate} if for any vertex $v \in V$, there exists some vertex $u \in V$ such that $\{u,v\} \in E$ is an unqualified edge. 
\end{definition}

\section{Results}
Our first main result is the necessary and sufficient condition for all CDS instances such that the capacity is $1/2$, stated in Theorem \ref{thm:1/2rate}.

\begin{theorem}\label{thm:1/2rate}
The capacity of CDS is $1/2$ if and only if within any qualified component, there is no internal qualified edge in an unqualified path.
\end{theorem}

The proof of Theorem \ref{thm:1/2rate} is presented in Section \ref{sec:thm1}.
Here to illustrate the idea, we give two examples. For the first one, the half-rate feasibility condition is satisfied and rate $1/2$ is achievable.

\begin{figure}[h]
\begin{center}
\includegraphics[width= 4.5 in]{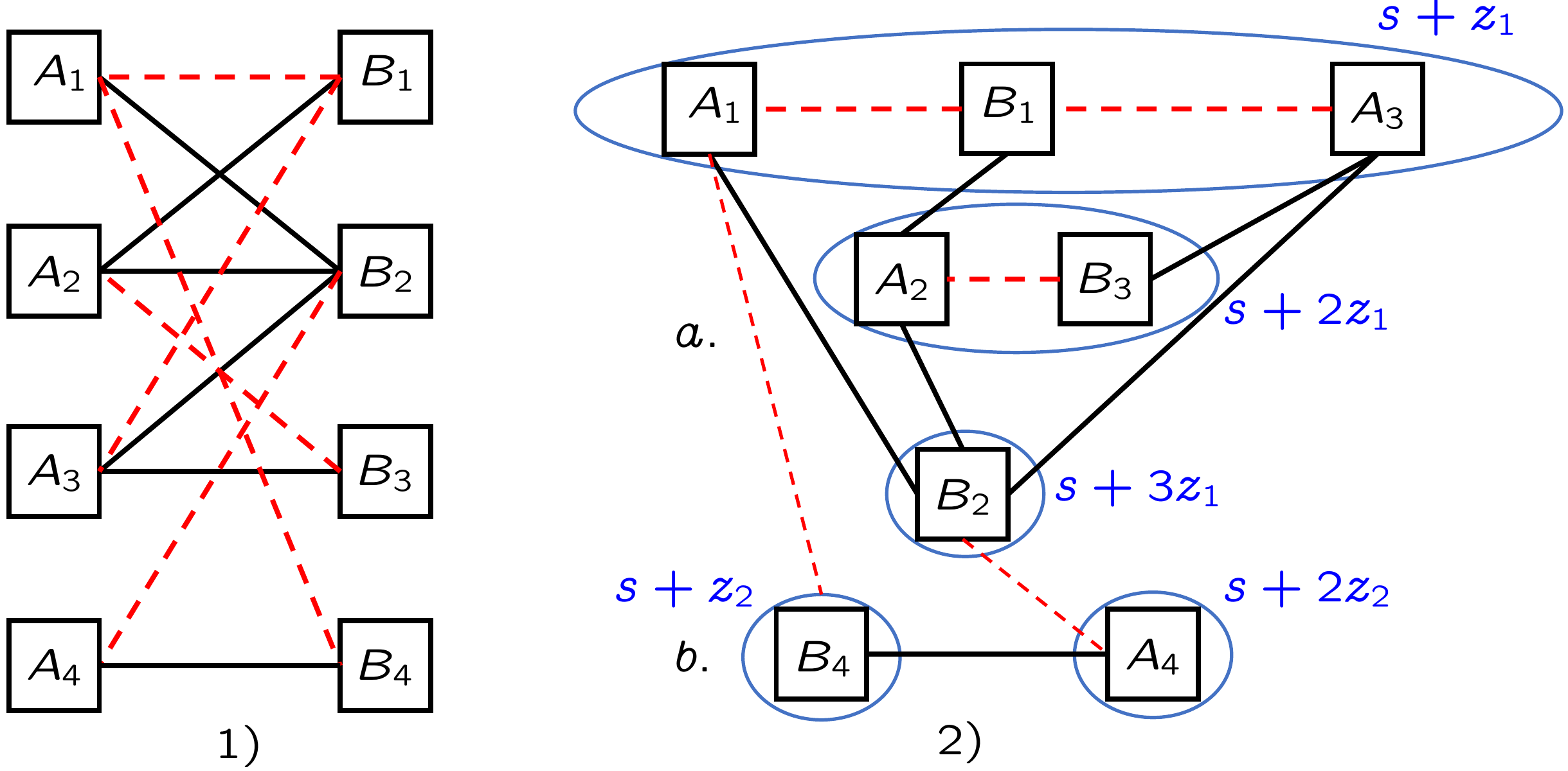}
\vspace{-0.15in}
\caption{\small 1). A CDS instance, described by its characteristic graph $G_f$. 2) The coding scheme that achieves rate $1/2$. The secret has 1 symbol $s$ from $\mathbb{F}_5$, the noise variable has 2 independent symbols $z_1, z_2$ from $\mathbb{F}_5$, and each signal has 1 symbol from $\mathbb{F}_5$. $G_f$ contains two qualified components (denoted by $a.$ and $b.$). %Any qualified component uses the same noise. Within a qualified component, any unqualified component uses the same signal. 
%From any qualified edge (solid black edge), we may recover the secret; from any unqualified edge (dashed red edge), we learn nothing about the secret.
}
\label{fig:ex1}
\end{center}
\end{figure}

\begin{example}\label{ex1}
Consider the CDS instance in Fig.~\ref{fig:ex1}.1, where the characteristic graph $G_f$ has two qualified components. Within qualified component $a.$, there are 3 unqualified paths and none of them has an internal qualified edge (see the blue circles in Fig.~\ref{fig:ex1}.2). Qualified component $b.$ only has one qualified edge and there is no unqualified path. Therefore, the half-rate feasibility condition in Theorem \ref{thm:1/2rate} is satisfied and the scheme that achieves rate $1/2$ is shown in Fig.~\ref{fig:ex1}.2. 

For the scheme, every vertex in a qualified component uses the same noise variable and different qualified components use independent noise variables (e.g., qualified components $a.$ and $b.$ use $z_1$ and $z_2$, respectively). Within a qualified component, we consider each unqualified component (a maximal set of vertices where any two vertices are connected by an unqualified path) sequentially, and assign each vertex in the unqualified component a linearly independent combination of the secret and noise (e.g., the 3 unqualified components in $a.$ are assigned $s+z_1, s+2z_1, s+3z_1$ respectively). Note that a vertex that is not connected to any unqualified edge (within a qualified component) is a (trivial) unqualified component (e.g., vertex $B_4$ in qualified component $b.$). 

The correctness constraint (\ref{dec}) holds because 1) any qualified edge belongs to a qualified component (e.g., $\{A_2, B_2\}$), 2) the two vertices belong to different unqualified components (note that there is no internal qualified edge, e.g., consider $A_2, B_2$), and 3) any distinct unqualified components are assigned a linearly independent combination of secrete and noise, from which the secret can be successfully recovered (e.g., $A_2 = s+2z_1, B_2 = s+3z_1$). We show that the security constraint (\ref{sec}) is guaranteed as well. There are two cases. First, for unqualified edges within a qualified component (e.g., $\{B_1, A_3\}$), they belong to the same unqualified component so that the same signal is assigned and no information about the secret is revealed (e.g., $B_1 = A_3 = s+z_1$). Second, for unqualified edges across two qualified components (e.g., $\{B_2, A_4\}$), different noise variables are used so that again nothing about the secret is leaked (e.g., $B_2 = s+3z_1, A_4 = s+2z_2$). 
\end{example}

\begin{figure}[h]
\begin{center}
\includegraphics[width= 4 in]{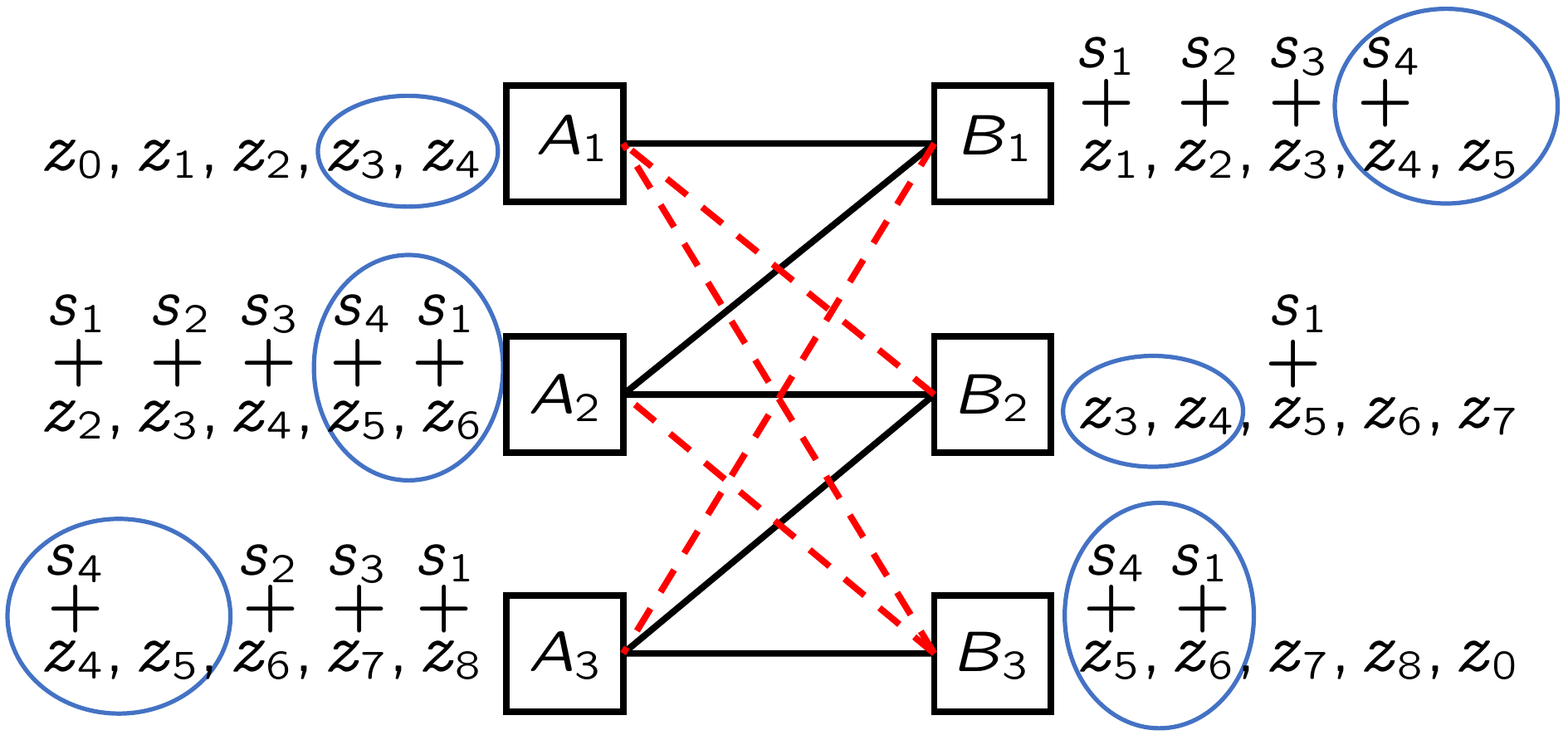}
\vspace{-0.15in}
\caption{\small A CDS instance that has an internal qualified edge $\{B_2, A_2\}$ in an unqualified path $(B_2, A_1, B_3, A_2)$ within a qualified component $G_f$, and the achievable scheme of rate $2/5$. The secret has $L = 4$ bits, $s_1, s_2, s_3, s_4$, the noise has $L_Z = 9$ independent uniform bits, $z_0, z_1, \cdots, z_8$, and each signal has $N=5$ bits. The rate achieved is $R = L/(2N) = 2/5$.
}
\label{fig:ex2}
\end{center}
\end{figure}

%\vspace{-0.1in}
For the second example, the condition in Theorem \ref{thm:1/2rate} is violated such that rate $1/2$ is not achievable. We use the CDS instance in Fig.~\ref{fig:prob} as the second example (reproduced in Fig.~\ref{fig:ex2}).

\vspace{-0.1in}

\vspace{-0.35in}
\begin{example}\label{ex2}
Consider the CDS instance in Fig.~\ref{fig:ex2}, where the characteristic graph $G_f$ is a qualified component. The unqualified path $(B_2, A_1, B_3, A_2)$ contains an internal qualified edge $\{B_2, A_2\}$, so the half-rate feasibility condition in Theorem \ref{thm:1/2rate} is violated and rate $1/2$ is not achievable. An intuitive explanation by contradiction is as follows.

Suppose rate $1/2$ is achievable, then the size of each signal $A_x, B_y$ that is connected to a qualified edge must be $N = L$ symbols and the noise appeared in the signal has size $L$ symbols as well (see Lemma \ref{lemma:size} in Section \ref{sec:thm11}). For any qualified edge, the noise variables for the two signals must be the same to ensure that the secret can be decoded, i.e., the noise space must fully overlap (see Lemma \ref{lemma:noise1} for a proof). For example, in Fig.~\ref{fig:ex2} $A_2, B_2$ must use the same noise. Then by sub-modularity, full noise alignment must hold for any qualified component, i.e., all signals in a qualified component must use the same noise variables (see Lemma \ref{lemma:noise2}). For example, in Fig.~\ref{fig:ex2} $A_1, A_2, A_3, B_1, B_2, B_3$ must use the same noise. Next, consider any unqualified edge, given that the noise space is fully overlapped, the signal space must fully overlap to avoid leaking information about the secret (see Lemma \ref{lemma:signal1}). For example, $B_2$ must be equal to $A_1$ in Fig.~\ref{fig:ex2}. Similarly by sub-modularity, for any unqualified path within a qualified component, the signal spaces must fully overlap (see Lemma \ref{lemma:signal2}). %in order to achieve rate $1/2$. 
For example, in Fig.~\ref{fig:ex2} we must have $B_2 = A_1 = B_3 = A_2$ for the unqualified path $(B_2, A_1, B_3, A_2)$. Finally, the presence of an internal qualified edge $\{B_2, A_2\}$ results in a contradiction, because $B_2 = A_2$ and $B_2$ is independent of the secret so that the edge $\{B_2, A_2\}$ cannot be qualified.
\end{example}

Note that rate $1/2$ is the highest for all non-degenerate settings as each vertex $v$ has at least one unqualified edge and the size of $v$ cannot be smaller than the secret size, i.e., $N \geq L$ and $R = L/(2N) \leq 1/2$. As the half-rate feasibility condition is fully settled, we proceed to scenarios where rate $1/2$ is not achievable. Interestingly, the simplest such instance is that in Fig.~\ref{fig:ex2}. This 6-node CDS instance is the simplest in the sense that for any 5-node non-degenerate CDS instance, half-rate feasibility condition is satisfied (easy to verify). 
Our second main result is the linear capacity characterization of the CDS instance in Fig.~\ref{fig:ex2}, stated in Theorem \ref{thm:2/5rate}.

%Explain simplest here

\begin{theorem}\label{thm:2/5rate}
The linear capacity of the CDS instance shown in Fig.~\ref{fig:ex2} is $2/5$.
\end{theorem}

The achievable scheme is shown in Fig.~\ref{fig:ex2}, where the secret has $L=4$ bits, $S = (s_1, s_2, s_3, s_4)$, and the noise has $L_Z = 9$ independent uniform bits, $Z = (z_0, z_1,\cdots, z_8)$. Each signal has $N = 5$ bits and is shown in Fig.~\ref{fig:ex2}. Note that along the qualified path $(A_1, B_1, A_2, B_2, A_3, B_3)$, every two connected vertices share 4 noise bits in a consecutive manner, i.e., $A_1$ uses $z_0, z_1, z_2, z_3, z_4$, and $B_1$ uses $z_1, z_2, z_3, z_4, z_5$ etc. 
The secret bits are assigned such that for any unqualified edge, the same noise bits are combined with the same secret bits (e.g., see the blue circles with the same shape in Fig.~\ref{fig:ex2}. For the unqualified edge $\{B_1, A_3\}$, both vertices use $z_4, z_5$ so that the same signal bits $s_4+z_4, z_5$ are present).

The rate achieved is $R = L/(2N) = 2/5$. Both correctness and security constraints are easy to verify. For example, consider the qualified edge $\{B_1, A_2\}$. Considering the part of the signal that uses the same noise $z_2, z_3, z_4, z_5$, we may recover $(s_1+s_2, s_2+s_3, s_3+s_4, s_4)$, from which we can decode $S = (s_1, s_2, s_3, s_4)$. %Other cases are similar. 
Consider the unqualified edge $\{A_2, B_3\}$. As distinct independent noise bits $z_2, z_3, z_4, z_7, z_8, z_0$ will not reveal anything and the common noise bits $z_5, z_6$ carry the same secret bits, security is guaranteed. 

The converse proof for all linear schemes is presented in Section \ref{sec:thm2}. We give an intuitive explanation of the idea here. A finer argument of the contradiction in Example \ref{ex2} is required. For this explanation, let us assume the noise space of any two vertices from a qualified edge share exactly $L$ dimensions in common (relaxation of this assumption is deferred to the full proof in Section \ref{sec:thm2}). That is, the noise spaces of $A_1$ and $B_1$ share $L$ dimensions (this space is denoted as $\gamma_1$), and $B_1$ and $A_2$ share $L$ dimensions (denote this space as $\gamma_2$). Now how many dimensions do $A_1, B_1, A_2$ have in common? $\gamma_1$ and $\gamma_2$ are two subspaces of the noise space of $B_1$ such that $\dim(\gamma_1 \cap \gamma_2) \geq \dim(\gamma_1) + \dim(\gamma_2) - N = 2L - N$. Proceeding with this argument along the qualified path $(A_1, B_1, A_2, B_2, A_3, B_3)$, we find that the noise spaces of $A_1, B_1, A_2, B_2, A_3, B_3$ must share $5L - 4N$ dimensions. We argue that such a common overlap cannot exist, so $5L - 4N \leq 0$ and $R_{linear} = L/(2N) \leq 2/5$. To set up the proof by contradiction, let us assume that all 6 noise spaces share a common dimension (denoted as $\gamma$). As the path $(B_2, A_1, B_3, A_2)$ is unqualified, the signal space of $\gamma$ must fully overlap as otherwise information about the secret will be revealed. This means that in the noise overlap of $\{A_2, B_2\}$, some signal is overlapped and does not contribute useful information of the secret. As the noise space of $A_2, B_2$ shares exactly $L$ dimensions and in the overlap $\gamma$ is useless, we cannot decode the $L$-symbol secret from $\{A_2, B_2\}$, arriving at the contradiction that $\{A_2, B_2\}$ is a qualified edge.
The intersections of more than 2 spaces have no correspondence to entropy terms such that the above linear argument may not hold in the information theoretic sense (i.e., non-linear codes might achieve a higher rate).
Note that the achievable scheme in Fig.~\ref{fig:ex2} is designed following the overlap insights provided by the linear converse idea.

%Explain converse
%{Achievability}

\section{Proof of Theorem \ref{thm:1/2rate}}\label{sec:thm1}

\subsection{Only if part}\label{sec:thm11}
Consider any non-degenerate CDS instance, described by the characteristic graph $G_f(V,E)$. We show that if the half-rate feasibility condition in Theorem \ref{thm:1/2rate} is violated, then rate $1/2$ is not achievable. To set up the proof by contradiction, let us assume that $R = L/2N = 1/2$ is achievable, i.e., $N=L$.
%\begin{eqnarray}
%N = L. \label{eq:nl}
%\end{eqnarray}
As a result, each signal that is connected to a qualified edge and the noise used in such a signal must have entropy $L$. This result is stated in Lemma \ref{lemma:size}.

\begin{lemma}[Signal and Noise Size] \label{lemma:size}
When $R = 1/2$, for any signal $v \in V$ such that there exists $u \in V$ such that $\{v,u\}$ is a qualified edge, we have
\begin{eqnarray}
H(v) = H(v | S) = L. \label{eq:size}
\end{eqnarray}
\end{lemma}

{\it Proof:}
First, consider the ``$\leq$'' direction. %We present the proof for $A_x$, and the proof for $B_x$ follows similarly.
\begin{eqnarray}
H(v | S) \leq H(v) \leq N = L.%\overset{(\ref{eq:nl})}{=} L
\end{eqnarray}
Second, consider the ``$\geq$'' direction. %We present the proof for $A_x$, and the proof for $B_x$ follows similarly. 
As the CDS instance is non-degenerate, for any vertex $w$, there exists a vertex $w'$ such that $\{w, w'\}$ is unqualified. From the security constraint (\ref{sec}), we have
\begin{eqnarray}
I(w, w'; S) = 0 &\Rightarrow& I(w; S) = 0 \\
\mbox{($w$ can by any vertex)} &\Rightarrow& I(v; S) = I(u;S) = 0. \label{eq:usind}
%\\
%&\Rightarrow& H(w | S) = H(w) 
\end{eqnarray}
Consider now the qualified edge $\{v,u\}$. From the correctness constraint (\ref{dec}), we have
\begin{eqnarray}
H(S | v,u) = 0 &\Rightarrow& L \overset{(\ref{sz_ind})}{=} H(S) = I(v,u; S) \overset{(\ref{eq:usind})}{=} I(v; S|u) \leq H(v)  \overset{(\ref{eq:usind})}{=} H(v|S). %\\
%&\overset{}{\Rightarrow}& H(S) \leq H(v) \overset{(\ref{eq:usind})}{=} H(v|S) 
\end{eqnarray}
The proof is thus complete.

\hfill\QED

Next, we consider any qualified edge and show that the noise appeared in both end vertices of the qualified edge has joint entropy $L$, same as the entropy of the noise appeared in each vertex by itself. In other words, the noise must fully align.

\begin{lemma}[Noise Alignment for Qualified Edge] \label{lemma:noise1}
When $R = 1/2$, for any qualified edge $\{v,u\}$, we have
\begin{eqnarray}
H(v,u | S) = L. \label{eq:edge}
\end{eqnarray}
\end{lemma}

{\it Proof:} On the one hand, we have
\begin{eqnarray}
H(v, u | S) &=& H(v, u, S) - H(S) \\
&\overset{(\ref{dec})}{=}& H(v,u) - H(S) \\
&\overset{(\ref{sz_ind})}{\leq}& H(v) + H(u) - L \\
&\overset{(\ref{eq:size})}{=}& L + L - L = L.
\end{eqnarray}
On the other hand, we have
\begin{eqnarray}
H(v, u | S) \geq H(v|S) &\overset{(\ref{eq:size})}{=}& L.
\end{eqnarray}
The proof is now complete.

\hfill\QED

In the following lemma, we generalize the noise alignment phenomenon from qualified edges to (any induced subgraph of) qualified components.
\begin{lemma}[Noise Alignment for Qualified Component] \label{lemma:noise2}
When $R = 1/2$, for any qualified component $Q$ with vertex set $V_Q \subset V$, we have
\begin{eqnarray}
\forall V_{q} \subset V_Q, ~H(V_{q} | S) = H(V_Q | S) = L. \label{eq:component}
\end{eqnarray}
\end{lemma}

{\it Proof:} We first prove the ``$\geq$'' direction.
\begin{eqnarray}
 H(V_Q | S) \geq H(V_{q} | S) &\geq& H(v | S) ~~\mbox{for any $v \in V_q$}  \\
 &\overset{(\ref{eq:size})}{=}& L. \label{eq:b1}
\end{eqnarray}
Second, we prove the ``$\leq$'' direction and complete the proof. Denote $V_Q = \{v_1, v_2, \cdots, v_Q\}$. Start with any qualified edge $\{v_{i_1}, v_{i_2}\}, i_1, i_2 \in \{1,2, \cdots, Q\}$ in the qualified component $Q$. As $Q$ is a qualified component, there must exist a vertex $v_{i_3} \in V_Q$ and a vertex from $v_{i_1}, v_{i_2}$ (suppose it is $v_{i_2}$ without loss of generality) such that $\{v_{i_2}, v_{i_3}\}$ is a qualified edge. From the sub-modularity property of entropy functions, we have
\begin{eqnarray}
 H(v_{i_1}, v_{i_2} | S) + H(v_{i_2}, v_{i_3} | S) &\geq& H(v_{i_1}, v_{i_2}, v_{i_3} |S) + H(v_{i_2} | S) \\
\overset{(\ref{eq:size}) (\ref{eq:edge})}{\Longrightarrow} ~~~~~~~~~~~~~~~~~~~~~~~~~~~~~~ L + L &\geq& H(v_{i_1}, v_{i_2}, v_{i_3} |S) + L\\
\overset{(\ref{eq:b1})}{\Rightarrow}~~~~~~~~~~~~~~~~~~ H(v_{i_1}, v_{i_2}, v_{i_3} |S) &=& L.
\end{eqnarray}
Then similarly, as $Q$ is a qualified component, there must exist a vertex $v_{i_4} \in V_Q$ such that $\{v, v_{i_4}\}$ is a qualified edge, where $v$ is one vertex from $v_{i_1}, v_{i_2}, v_{i_3}$. With a similar proof as above, we have
\begin{eqnarray}
H(v_{i_1}, v_{i_2}, v_{i_3}, v_{i_4} |S) = L \Rightarrow \cdots \Rightarrow H(V_Q | S) = L \Rightarrow H(V_q | S) \leq H(V_Q | S) = L.
\end{eqnarray}
%As $V_Q$ is the vertex set of a qualified component, any two vertices $v_i, v_j \in V_Q$ are connected. 

\hfill\QED

We now proceed to the signal alignment phenomenon. We show that within a qualified component, any two vertices $v,u$ that form an unqualified edge must produce exactly the same signal, i.e., the joint entropy of $v, u$ is $L$, which is the same as that of any individual $v$ or $u$ by itself.

\begin{lemma}[Signal Alignment for Unqualified Edge within Qualified Component] \label{lemma:signal1}
When $R = 1/2$, for any unqualified edge $\{v,u\}$ that is within a qualified component $Q$, we have
\begin{eqnarray}
H(v,u) = L. \label{eq:unedge}
\end{eqnarray}
\end{lemma}

{\it Proof:} Note that both end vertices of the unqualified edge $\{v,u\}$ belong to the vertex set of the qualified component $Q$. Combining the security constraint (\ref{sec}) and (\ref{eq:component}), we have
\begin{eqnarray}
H(v,u) \overset{(\ref{sec})}{=} H(v,u | S) \overset{(\ref{eq:component})}{=} L.
\end{eqnarray}

\hfill \QED

In the following lemma, we generalize the signal alignment phenomenon from unqualified edges to unqualified paths.

\begin{lemma}[Signal Alignment for Unqualified Path within Qualified Component] \label{lemma:signal2}
When $R = 1/2$, for any unqualified path within a qualified component $Q$, $(\{v_1, v_2\}, \{v_2, v_3\}, \cdots, \{v_{P-1}, v_P\})$, we have
\begin{eqnarray}
H(v_1, v_P) \leq L. \label{eq:unpath}
\end{eqnarray}
\end{lemma}

{\it Proof:} Equipped with what has been established, the proof follows from a simple recursive application of the sub-modularity property of entropy functions.
\begin{eqnarray}
H(v_1, v_2) + H(v_2, v_3) + \cdots H(v_{P-1}, v_P) &\geq& H(v_1, v_2, \cdots, v_P) + H(v_2) + H(v_3) + \cdots + H(v_{P-1}) \notag\\
\overset{(\ref{eq:unedge}) (\ref{eq:size})}{\Rightarrow} ~~~~~~~~~~ (P-1) L &\geq& H(v_1, v_P) + (P-2)L ~~~~~~ \Rightarrow H(v_1, v_P) \leq L.
\end{eqnarray}

%The other ``$\geq$'' direction is also simple.
%\begin{eqnarray}
%H(v_1, v_P) \geq H(v_1) \overset{(\ref{eq:size})}{=} L
%\end{eqnarray} 
\hfill \QED

After establishing the above lemmas, we are ready to present where is the contradiction. As the half-rate feasibility condition is violated, there must exist an internal qualified edge (denoted as $\{v_1, v_P\}$) in an unqualified path $(\{v_1, v_2\}, \{v_2, v_3\}, \cdots, \{v_{P-1}, v_P\})$ and the unqualified path is within a qualified component $Q$. From the correctness constraint (\ref{sec}) of the qualified edge $\{v_1, v_P\}$, we have %(i.e., $v_1, v_2, \cdots, v_P$)
\begin{eqnarray}
L \overset{(\ref{eq:unpath})}{\geq} H(v_1, v_P) \overset{(\ref{dec})}{=} H(v_1, v_P, S) = H(S) + H(v_1, v_P | S) \overset{(\ref{sz_ind})(\ref{eq:edge})}{=}  L + L.
\end{eqnarray}
So $L \geq 2L$, and we have arrived at the contradiction.
The proof of the only if part is thus complete.

\begin{remark}
The above proof is based on assuming that $R = 1/2$ and then arguing by contradiction. We may bound the terms appeared more carefully and obtain a stronger bound on $R$, $R \leq \overline{R}$, where $\overline{R}$ is strictly smaller than $1/2$, i.e., $\overline{R} = 1/2 - \delta$ for a positive constant $\delta$. We do not choose to provide a concrete value of $\overline{R}$, because 1) such a proof will be more lengthy and ideas are less clear, 2) and the bound produced by this procedure may not be the tightest bound possible from all sub-modularity constraints (see Remark \ref{remark:best}).
\end{remark}

\subsection{If part}
We show that if the half-rate feasibility condition in Theorem \ref{thm:1/2rate} is satisfied, then the CDS capacity is 1/2. We first prove that $R \leq 1/2$ and then show that $R = 1/2$ is achievable.

The proof of $R \leq 1/2$ is as follows. Consider any CDS instance that contains at least one qualified edge $\{v,u\}$; otherwise all edges are unqualified, the problem is meaningless as the secret is never disclosed. Further, the CDS instance is non-degenerate, so there exists an unqualified edge $\{u,w\}$. From the security constraint (\ref{sec}), we have
\begin{eqnarray}
I(u,w ; S) = 0 \Rightarrow I(u;S) = 0. \label{eq:i1}
\end{eqnarray}
From the correctness constraint (\ref{dec}), we have
\begin{eqnarray}
&& L \overset{(\ref{sz_ind})}{=} H(S) \overset{(\ref{dec})}{=} I(S; v,u) \overset{(\ref{eq:i1})}{=} I(S; v | u) \leq H(v) \leq N \\
&\Rightarrow& R = L/(2N) \leq 1/2.
\end{eqnarray}

We now present the coding scheme that achieves rate $1/2$. The scheme is a generalization of that presented in Example \ref{ex1}.

Consider any non-degenerate CDS instance, described by the characteristic graph $G_f(V,E)$. Suppose $G_f(V,E)$ has $M$ qualified components. A single vertex that is not connected to any qualified edge is a (trivial) qualified component. Suppose within the $m^{th}, m \in \{1,2,\cdots, M\}$ qualified component, there are $U_m$ unqualified components. %Denote $U_{\max} = {\max}_m U_m$. 
Choose $p$ as a prime number that is no fewer than $\max(U_1, U_2, \cdots, U_M)$. The secret $S$ contains $L = 1$ symbol from the finite field $\mathbb{F}_p$, denoted as $S = (s)$ and the noise $Z$ contains $L_Z = M$ symbols from $\mathbb{F}_p$, denoted as $Z = (z_1, z_2, \cdots, z_M)$. Note that $z_1, \cdots, z_M$ are i.i.d. uniform symbols over $\mathbb{F}_p$.

The signals are assigned as follows. Consider the $m^{th}$ qualified component $Q_m$. We set
\begin{eqnarray}
\mbox{any signal $v$ in the $i^{th}, i \in \{1,2,\cdots,U_m\}$ unqualified component within $Q_m$ as} ~ s + i z_m. \label{eq:scheme}
\end{eqnarray}

To complete the proof of the achievable scheme, we show that the scheme is both correct and secure. Consider the correctness constraint (\ref{dec}) first. A qualified edge must belong to one qualified component. As the half-rate feasibility condition in Theorem \ref{thm:1/2rate} is satisfied, there is no internal qualified edge, i.e., any qualified edge must belong to different unqualified components within a qualified component. Consider any qualified edge $\{v,u\}$ that is from qualified component $Q_m$ and within $Q_m$, suppose $v$ belongs to the $i^{th}$ unqualified component and $u$ belongs to the $j^{th}$ unqualified component. Note that $j$ is not equal to $i$. From (\ref{eq:scheme}), we have
\begin{eqnarray}
&& v = s + i z_m, u = s + j z_m \\
&\Rightarrow& H(S | v,u)  = H(s | s + i z_m, s + j z_m) \overset{j \neq i}{=} H(s | s, z_m) = 0
\end{eqnarray}
so that the scheme is always correct.

Next consider the security constraint (\ref{sec}). Consider any unqualified edge $\{v,u\}$. We have the following two cases.
\begin{enumerate}
\item $\{v,u\}$ is from the same qualified component, say $Q_m$. Note that any unqualified edge must belong to the same unqualified component within $Q_m$, say the $i^{th}$ unqualified component. From (\ref{eq:scheme}), we have
\begin{eqnarray}
&& v = u = s + i z_m \label{eq:zz1} \\
&\Rightarrow& H(S | v,u)  = H(s | s + i z_m) = H(s, s+i z_m) - H(s+iz_m) = 1 = H(S)
\end{eqnarray}
so that security is guaranteed.

\item $\{v,u\}$ is from different qualified components. Suppose $v$ is from $Q_m$ and $u$ is from $Q_{m'}$, where $m \neq m'$. Further assume that $v$ belongs to the $i^{th}$ unqualified component in $Q_m$, and $u$ belongs to the $j^{th}$ unqualified component in $Q_{m'}$. From (\ref{eq:scheme}), we have
\begin{eqnarray}
&& v = s + i z_m, u = s + j z_{m'} \\
&\Rightarrow& H(S | v,u)  = H(s | s + i z_m, s + j z_{m'}) \\
&&~~~~~~~~~~~~ = H(s, s + i z_m, s + j z_{m'}) - H(s + i z_m, s + j z_{m'}) \\
&&~~~~~~~~~~~~ = H(s, z_m, z_{m'}) - H(s + i z_m, s + j z_{m'})  \\
&&~~~~~~~~~~~~ \geq H(s, z_m, z_{m'}) - 2 = 1 = H(S) \label{eq:zz2}
\end{eqnarray}
so that $H(S | v,u) = H(S)$ and security is guaranteed.
\end{enumerate}

\subsection*{Randomness Cost Reduction}
The above scheme uses $M$ noise symbols in total. We show that $2$ noise symbols are sufficient, i.e., we save $M-2$ noise symbols and the randomness rate is improved from $R_Z = 1/M$ to $R_Z = 1/2$. This reduction is made possible by the following simple observation - each unqualified edge only involves two vertices and each vertex only contains 1 noise symbol, so we only need to guarantee these two noise symbols appeared (if different) are linearly independent and for this purpose, two base noise symbols are sufficient as all other noise symbols can be generic linear combinations of these two noise symbols. The detailed proof is presented next.

Choose $p$ as a prime number such that $p > \max(U_1, U_2, \cdots, U_M, M-2)$. The remaining proof is the same as that above, except that $z_1, z_2. \cdots, z_M$ are linear combinations of two base independent uniform symbols $z_1, z_2$ (instead of being mutually independent).
\begin{eqnarray}
z_3 = z_1 + z_2, ~z_4 = z_1 + 2z_2, \cdots, ~z_M = z_1 + (M-2)z_2. \label{eq:newz}
\end{eqnarray}
The correctness constraint is not influenced as only the noise assignment is changed. The security constraint continues to hold as we may easily verify that every step in (\ref{eq:zz1}) - (\ref{eq:zz2}) goes through after we set (\ref{eq:newz}).

\begin{remark}
While the characteristic graph of a CDS instance is bipartite, a closer inspection of the proof of both the only if part and the if part reveals that the bipartite property is not used in the proof. Therefore Theorem \ref{thm:1/2rate} holds also for non-bipartite characteristic graphs.
\end{remark}

\section{Proof of Theorem \ref{thm:2/5rate}: Linear Converse} \label{sec:thm2}
We show that for the CDS instance in Fig.~\ref{fig:ex2}, the rate for all linear schemes cannot be higher than $2/5$. For a linear scheme, the signal $v$ is a linear function of the secret $S \in \mathbb{F}_p^{L \times 1}$ and the noise $Z \in \mathbb{F}_p^{L_Z \times 1}$. All secret and noise symbols are i.i.d. and uniform.
\begin{eqnarray}
v = {\bf F}_{v} S + {\bf H}_v Z  
\end{eqnarray}
where ${\bf F}_v$ is an $N\times L$ matrix over $\mathbb{F}_p$, and ${\bf H}_v$ is an $N \times L_Z$ matrix over $\mathbb{F}_p$.

We first establish two general properties that hold for all linear schemes. The first property states that for any qualified edge, the overlap of the noise spaces cannot be fewer than $L$ dimensions. This property is stated in Lemma \ref{lemma:noverlap}.

\begin{lemma}[Noise Alignment] \label{lemma:noverlap}
For any linear scheme and for any qualified edge $\{v,u\}$, we have
\begin{eqnarray}
%L \leq \dim( \mbox{rowspan}({\bf H}_v) \cap \mbox{rowspan}({\bf H}_u) ) \leq N
\dim( \mbox{rowspan}({\bf H}_v) \cap \mbox{rowspan}({\bf H}_u) ) \geq L. \label{eq:nalign}
\end{eqnarray}
\end{lemma}

{\it Proof:} %As the dimension of ${\bf H}_v, {\bf H}_u$ is $N \times L_Z$, it follows that $\dim( \mbox{rowspan}({\bf H}_v) \cap \mbox{rowspan}({\bf H}_u) ) \leq \dim({\bf H}_u) \leq N$ and we only need to prove that $\dim( \mbox{rowspan}({\bf H}_v) \cap \mbox{rowspan}({\bf H}_u) ) \geq L$.
For any non-degenerate setting, we know from (\ref{eq:i1}) that any vertex must be independent of the secret.
\begin{eqnarray} 
0 = I(S; v) = I(S; {\bf F}_{v} S + {\bf H}_v Z) &\Rightarrow& 0 = I(S; {\bf F}_{v}(\mathcal{J}, :) S + {\bf H}_v(\mathcal{J}, :) Z) \label{eq:cover}
\end{eqnarray}
where $\mathcal{J} \subset \{1,2,\cdots,N\}$ and for a matrix $A$, we use $A(\mathcal{J}, :)$ to denote the sub-matrix of $A$ formed by rows in the index set $\mathcal{J}$.
In words, (\ref{eq:cover}) means that for linear schemes the secret space must be fully covered by the noise space.

Denote $\dim( \mbox{rowspan}({\bf H}_v) \cap \mbox{rowspan}({\bf H}_u) )$ by $\alpha$. As ${\bf H}_v$ and ${\bf H}_u$ overlap in $\alpha$ dimensions, we may assume without loss of generality (by a change of basis operation) that the first $\alpha$ rows of ${\bf H}_u$ and ${\bf H}_v$ are the same, i.e., ${\bf H}_v (1:\alpha, :) = {\bf H}_u (1:\alpha, :) \triangleq {\bf H}_\alpha$. Further, we have that
\begin{eqnarray}
\mbox{The row vectors of ${\bf H}_\alpha$, ${\bf H}_v (\alpha+1:N, :)$  and ${\bf H}_u (\alpha+1:N, :)$ are linearly independent.}  \label{eq:in1}
\end{eqnarray}
To simplify the notation, we define
\begin{eqnarray}
{\bf F}_{v}(1:\alpha, :) \triangleq {\bf F}_{v_1}, && {\bf F}_{v}(\alpha+1, N, :) \triangleq {\bf F}_{v_2}, \\
&& {\bf H}_{v}(\alpha+1, N, :) \triangleq {\bf H}_{v_2}.
\end{eqnarray}

For the qualified edge $\{v, u\}$, the correctness constraint (\ref{dec}) requires that
\begin{eqnarray}
L \overset{(\ref{sz_ind})}{=} 
H(S) &=& I(S; v, u) \\
&=& I(S ; {\bf F}_{v_1} S + {\bf H}_\alpha Z, {\bf F}_{u_1}S + {\bf H}_\alpha Z, {\bf F}_{v_2} S + {\bf H}_{v_2} Z, {\bf F}_{u_2} S + {\bf H}_{u_2} Z) \\
&=& I(S ; ({\bf F}_{v_1} - {\bf F}_{u1})S , {\bf F}_{u_1}S + {\bf H}_\alpha Z, {\bf F}_{v_2} S + {\bf H}_{v_2} Z, {\bf F}_{u_2} S + {\bf H}_{u_2} Z) \\
&=& I(S ; {\bf F}_{u_1}S + {\bf H}_\alpha Z, {\bf F}_{v_2} S + {\bf H}_{v_2} Z, {\bf F}_{u_2} S + {\bf H}_{u_2} Z) \notag\\
&& +~ I(S ; ({\bf F}_{v_1} - {\bf F}_{u_1})S | {\bf F}_{u_1}S + {\bf H}_\alpha Z, {\bf F}_{v_2} S + {\bf H}_{v_2} Z, {\bf F}_{u_2} S + {\bf H}_{u_2} Z) \label{eq:k0} \\
&\leq& H(({\bf F}_{v_1} - {\bf F}_{u_1})S) \label{eq:k1} \\
&\leq& \alpha
\end{eqnarray}
where (\ref{eq:k1}) follows from the property that the first term of (\ref{eq:k0}) is zero (proved in the following), and the last step follows from the fact that $({\bf F}_{v_1} - {\bf F}_{u_1})S$ has at most $\alpha$ symbols.

To complete the proof of $\alpha \geq L$, we show that $I(S ; {\bf F}_{u_1}S + {\bf H}_\alpha Z, {\bf F}_{v_2} S + {\bf H}_{v_2} Z, {\bf F}_{u_2} S + {\bf H}_{u_2}Z) = 0$.
\begin{eqnarray}
&& I(S ; {\bf F}_{u_1}S + {\bf H}_\alpha Z, {\bf F}_{v_2} S + {\bf H}_{v_2} Z, {\bf F}_{u_2} S + {\bf H}_{u_2}Z) \notag\\
&=& H({\bf F}_{u_1}S + {\bf H}_\alpha Z, {\bf F}_{v_2} S + {\bf H}_{v_2} Z, {\bf F}_{u_2} S + {\bf H}_{u_2}Z) \notag\\
&& -~ H({\bf F}_{u_1}S + {\bf H}_\alpha Z, {\bf F}_{v_2} S + {\bf H}_{v_2} Z, {\bf F}_{u_2} S + {\bf H}_{u_2}Z | S) \\
&\leq& H({\bf F}_{u_1}S + {\bf H}_\alpha Z) + H({\bf F}_{v_2} S + {\bf H}_{v_2} Z) + H({\bf F}_{u_2} S + {\bf H}_{u_2}Z) \notag\\
&& -~ H({\bf H}_\alpha Z, {\bf H}_{v_2} Z, {\bf H}_{u_2} Z) \\
&\overset{(\ref{eq:cover}) (\ref{eq:in1})}{=}& H({\bf H}_\alpha Z) + H({\bf H}_{v_2} Z) + H({\bf H}_{u_2}Z) - H({\bf H}_\alpha Z) - H({\bf H}_{v_2} Z) - H({\bf H}_{u_2} Z) = 0.
\end{eqnarray}
As mutual information is non-negative, the proof of Lemma \ref{lemma:noverlap} is now complete.

\hfill\QED

The second property states that for any unqualified edge, within the noise overlapping space, the signal space must fully overlap. This property is stated in Lemma \ref{lemma:soverlap}.

\begin{lemma}[Signal Alignment]\label{lemma:soverlap}
For any linear scheme and for any unqualified edge $\{v,u\}$, we have
\begin{eqnarray}
\forall \mathcal{J} \subset \{1,2,\cdots,N\}, ~~{\bf H}_v(\mathcal{J}, :) = {\bf H}_u(\mathcal{J}, :) &\Rightarrow& {\bf F}_v(\mathcal{J}, :) = {\bf F}_u(\mathcal{J}, :). \label{eq:salign}
\end{eqnarray} 
\end{lemma}

{\it Proof:}
For the unqualified edge $\{v,u\}$, the security constraint (\ref{sec}) imposes that
\begin{eqnarray}
%&& 
0 = I(S; v,u) &=& I(S; {\bf F}_{v} S + {\bf H}_v Z, {\bf F}_{u} S + {\bf H}_u Z) \notag \\
%&\Rightarrow& 0 
&\geq& I(S; {\bf F}_{v}(\mathcal{J}, :)  S + {\bf H}_v(\mathcal{J}, :) Z, {\bf F}_{u}(\mathcal{J}, :) S + {\bf H}_u(\mathcal{J}, :) Z) \\
%&\Rightarrow& 0 
&\geq& I( S; ({\bf F}_{v}(\mathcal{J}, :) - {\bf F}_{u}(\mathcal{J}, :))  S + ( {\bf H}_v(\mathcal{J}, :) - {\bf H}_u(\mathcal{J}, :) ) Z). 
\end{eqnarray}
Now suppose. ${\bf H}_v(\mathcal{J}, :) = {\bf H}_u(\mathcal{J}, :)$. Plugging this condition into the equality above, we have
\begin{eqnarray}
0 \geq I( S; ({\bf F}_{v}(\mathcal{J}, :) - {\bf F}_{u}(\mathcal{J}, :))  S)  &\Rightarrow& {\bf F}_{v}(\mathcal{J}, :) = {\bf F}_{u}(\mathcal{J}, :)
\end{eqnarray}
and the proof is complete.

\hfill\QED

Equipped with the above two lemmas, we are ready to consider the CDS instance in Fig.~\ref{fig:ex2}. We first consider the qualified path $P = (\{A_1, B_1\}, \{B_1, A_2\}, \{A_2, B_2\}, \{B_2, A_3\}, \{A_3, B_3\})$ and see what is the dimension of the common overlap for the noise spaces of $A_1, B_1, A_2, B_2, A_3, B_3$. For any given linear scheme, we find the noise overlap of every qualified edge in $P$ and simplify the notation as follows.
\begin{eqnarray}
\dim( \mbox{rowspan}({\bf H}_{v}) \cap \mbox{rowspan}({\bf H}_{u}) ) \triangleq \alpha_{vu}, ~\mbox{e.g.,}~
\dim( \mbox{rowspan}({\bf H}_{A_1}) \cap \mbox{rowspan}({\bf H}_{B_1}) ) = \alpha_{A_1B_1}
%, \dim( \mbox{rowspan}({\bf H}_{B_1}) \cap \mbox{rowspan}({\bf H}_{A_2}) ) \triangleq \alpha_{B_1A_2} \\
%\dim( \mbox{rowspan}({\bf H}_{A_1}) \cap \mbox{rowspan}({\bf H}_{B_1}) ) \triangleq \alpha_{A_1B_1}, \dim( \mbox{rowspan}({\bf H}_{B_1}) \cap \mbox{rowspan}({\bf H}_{A_2}) ) \triangleq \alpha_{B_1A_2}\\
%\dim( \mbox{rowspan}({\bf H}_{A_1}) \cap \mbox{rowspan}({\bf H}_{B_1}) ) \triangleq \alpha_{A_1B_1}, \dim( \mbox{rowspan}({\bf H}_{B_1}) \cap \mbox{rowspan}({\bf H}_{A_2}) ) \triangleq \alpha_{B_1A_2}
\end{eqnarray}
and we identify 5 constants $\alpha_{A_1B_1}, \alpha_{B_1A_2}, \alpha_{A_2B_2}, \alpha_{B_2A_3}, \alpha_{A_3B_3}$.
Similarly, we denote
\begin{eqnarray}
\dim( \mbox{rowspan}({\bf H}_{v}) \cap \mbox{rowspan}({\bf H}_{u}) \cap \mbox{rowspan}({\bf H}_{w}) ) = \alpha_{vuw}, ~\mbox{etc.}
\end{eqnarray}
and wish to characterize $\alpha_{A_1B_1A_2B_2A_3B_3}$, i.e., the overlap of 6 noise spaces. Consider $\alpha_{A_1B_1A_2}$.
\begin{eqnarray}
\alpha_{A_1B_1A_2} &=& \dim\big( \mbox{rowspan}({\bf H}_{A_1}) \cap \mbox{rowspan}({\bf H}_{B_1}) \cap  \mbox{rowspan}({\bf H}_{A_2}) \big) \notag \\
&=&\dim\big( ( \mbox{rowspan}({\bf H}_{A_1}) \cap \mbox{rowspan}({\bf H}_{B_1}) ) \cap  ( \mbox{rowspan}({\bf H}_{B_1}) \cap \mbox{rowspan}({\bf H}_{A_2}) )  \big)\\
&\geq& \dim\big( ( \mbox{rowspan}({\bf H}_{A_1}) \cap \mbox{rowspan}({\bf H}_{B_1}) ) \big) + \dim\big( ( \mbox{rowspan}({\bf H}_{B_1}) \cap \mbox{rowspan}({\bf H}_{A_2}) ) \big) \notag\\
&& -~ \dim( \mbox{rowspan}({\bf H}_{B_1}) ) \label{eq:o1} \\
&=& \alpha_{A_1B_1} + \alpha_{B_1A_2} - N \label{eq:o2}
\end{eqnarray}
where (\ref{eq:o1}) follows from the fact that both the overlap of the row span of ${\bf H}_{A_1}$, ${\bf H}_{B_1}$ and the overlap of the row span of ${\bf H}_{B_1}$, ${\bf H}_{A_2}$ are subspaces of ${\bf H}_{B_1}$, and within a vector space of dimension $\alpha$, two subspaces of dimension $\alpha_1, \alpha_2$ must overlap in a space of dimension at least $\alpha_1 + \alpha_2 - \alpha$. (\ref{eq:o2}) is due to the fact that we may assume without loss of generality $\dim( \mbox{rowspan}({\bf H}_{B_1}) ) = N$, i.e., the noise space has full rank. This is argued as follows. Suppose the matrix ${\bf H}_{B_1}$ does not have full row rank, i.e., there exists a row of ${\bf H}_{B_1}$ that is a linear combination of other rows, say ${\bf H}_{B_1}(1,:)$. From (\ref{eq:cover}), we know that $B_1$ is independent of $S$, so the precoding vector of the secret ${\bf F}_{B_1}(1,:)$ must also be the same linear combination of other rows of ${\bf F}_{B_1}$. In other words, the first row of the signal $B_1$ is a deterministic function of the other rows of $B_1$ and contributes no entropy to $B_1$ (thus can be eliminated without loss). So we may only consider achievable schemes so that for any signal, the precoding matrix for the noise has full rank\footnote{The noise precoding matrix has size $N\times L_Z$, where $N \leq L_Z$. Note that if otherwise $N > L_Z$, then the rows of the noise cannot be linearly independent, and we have a similar situation where some row is a linear combination of other rows and we can follow the same line to argue that this row of signal is redundant.}, $N$.

We proceed similarly to the overlap of 4 noise spaces, $\alpha_{A_1B_1A_2B_2}$. Interpreting this overlap as the overlap of two spaces, i.e., the row span of ${\bf H}_{A_1}, {\bf H}_{B_1}, {\bf H}_{A_2}$ and the row space of ${\bf H}_{A_2}, {\bf H}_{B_2}$, within one space, i.e., the row space of ${\bf H}_{A_2}$, we have
\begin{eqnarray}
\alpha_{A_1B_1A_2B_2} &\geq& \alpha_{A_1B_1A_2} + \alpha_{A_2B_2} - N \\
&\overset{(\ref{eq:o2})}{\geq}& \alpha_{A_1B_1} + \alpha_{B_1A_2} + \alpha_{A_2B_2} - 2N \\
\mbox{Similarly,}~ \alpha_{A_1B_1A_2B_2A_3} &\geq& \alpha_{A_1B_1} + \alpha_{B_1A_2} + \alpha_{A_2B_2} + \alpha_{B_2A_3} - 3N \\
\alpha_{A_1B_1A_2B_2A_3B_3} &\geq& \alpha_{A_1B_1} + \alpha_{B_1A_2} + \alpha_{A_2B_2} + \alpha_{B_2A_3} + \alpha_{A_3B_3} - 4N \triangleq \alpha^* . \label{eq:as}
\end{eqnarray}
In other words, the 6 noise spaces overlap in a space of dimension at least $\alpha^*$ so that we may assume without loss of generality that the first $\alpha^*$ rows of the noise precoding matrix of $A_1, B_1, A_2, B_2, A_3, B_3$ are the same.
\begin{eqnarray}
{\bf H}_{A_1} (1:\alpha^*, :) = {\bf H}_{B_1} (1:\alpha^*, :) = {\bf H}_{A_2} (1:\alpha^*, :) = {\bf H}_{B_2} (1:\alpha^*, :) = {\bf H}_{A_3} (1:\alpha^*, :) = {\bf H}_{B_3} (1:\alpha^*, :).
\end{eqnarray} 

We next consider the unqualified path $P_u = (\{B_2, A_1\}, \{A_1, B_3\}, \{B_3, A_2\})$, where every vertex belongs to the qualified path $P$ considered above. The overlap of the 6 noise spaces must be a subspace of the overlap of the noise space of any unqualified edge. Applying Lemma \ref{lemma:soverlap}, i.e., (\ref{eq:salign}) to the 3 unqualified edges in $P_u$, we have
\begin{eqnarray}
{\bf H}_{B_2} (1:\alpha^*, :) = {\bf H}_{A_1} (1:\alpha^*, :) &\Rightarrow& {\bf F}_{B_2} (1:\alpha^*, :) = {\bf F}_{A_1} (1:\alpha^*, :) \\
{\bf H}_{A_1} (1:\alpha^*, :) = {\bf H}_{B_3} (1:\alpha^*, :) &\Rightarrow& {\bf F}_{A_1} (1:\alpha^*, :) = {\bf F}_{B_3} (1:\alpha^*, :) \\
{\bf H}_{B_3} (1:\alpha^*, :) = {\bf H}_{A_2} (1:\alpha^*, :) &\Rightarrow& {\bf F}_{B_3} (1:\alpha^*, :) = {\bf F}_{A_2} (1:\alpha^*, :) \\
\Rightarrow~~ {\bf H}_{B_2} (1:\alpha^*, :) = {\bf H}_{A_2} (1:\alpha^*, :), && {\bf F}_{B_2} (1:\alpha^*, :) = {\bf F}_{A_2} (1:\alpha^*, :). \label{eq:a2b2}
\end{eqnarray}

The final step is to consider the internal qualified edge $\{A_2,B_2\}$, where we have the noise and signal alignment constraint (\ref{eq:a2b2}). The correctness constraint (\ref{dec}) requires that
\begin{eqnarray}
L  \overset{(\ref{sz_ind})}{=} H(S) = I(S; A_2, B_2).
\end{eqnarray}
Following the proof of (\ref{eq:k1}), we have
\begin{eqnarray}
L &\leq& H( ({\bf F}_{A_2} (1:\alpha_{A_2B_2}, :) - {\bf F}_{B_2} (1:\alpha_{A_2B_2}, :) )S) \\
&\overset{(\ref{eq:a2b2})}{=}& H( ({\bf F}_{A_2} (\alpha^*+1:\alpha_{A_2B_2}, :) - {\bf F}_{B_2} (\alpha^*+1:\alpha_{A_2B_2}, :) )S) \\
&\leq& \alpha_{A_2B_2} - \alpha^* \\
&\overset{(\ref{eq:as})}{=}&\alpha_{A_2B_2} - ( \alpha_{A_1B_1} + \alpha_{B_1A_2} + \alpha_{A_2B_2} + \alpha_{B_2A_3} + \alpha_{A_3B_3} - 4N)\\
&=& 4N - (\alpha_{A_1B_1} + \alpha_{B_1A_2}  + \alpha_{B_2A_3} + \alpha_{A_3B_3}) \\
&\overset{(\ref{eq:nalign})}{\leq}& 4N - 4L \\
\Rightarrow ~~ R &=& L/(2N) ~\leq~ 4N/5 \times 1/(2N) = 2/5.
\end{eqnarray}
The linear converse proof is thus complete.

\begin{remark} \label{remark:best}
The information theoretic capacity of the CDS instance in Fig.~\ref{fig:ex2} is an interesting open problem, which might be challenging. While the linear capacity is characterized in Theorem \ref{thm:2/5rate} to be $2/5$, the best information theoretic converse with all Shannon type information inequalities \cite{Y_ITNC} (sub-modularity constraints) is $5/12$, found by computer programs \cite{XITIP, TPH}. Therefore, if the linear scheme of Theorem \ref{thm:2/5rate} is information theoretically optimal, then we need non-Shannon type information inequalities to establish the converse; if the best converse with only Shannon-type information inequalities is information theoretically optimal, then we need non-linear codes to achieve it. Therefore, for the CDS instance in Fig.~\ref{fig:ex2} with only 6 nodes and defined by only 8 variables (if only capacity is of concern, this can be reduced to 7 by eliminating the noise variable), either non-linear codes are necessary for achievability schemes or non-Shannon inequalities are necessary for converse arguments; further it is possible that both are required to establish the capacity.

For the best converse of $5/12$ with only Shannon-type information inequalities, we have not found a proof by hand. Through tightening the steps in the proof of Theorem \ref{thm:1/2rate} in a non-trivial manner (details omitted as no new insights emerge), we can obtain a converse bound of $3/7$, while a naive application of the inequalities in Theorem \ref{thm:1/2rate} induces a looser converse bound of $11/24$.
\end{remark}

%{\color{blue} \subsection{Information Theoretic Converse}}

\section{Conclusion}
The conditional disclosure of secrets problem is studied from an information theoretic capacity perspective. A noise and signal alignment approach is used to identify all best case scenarios where the capacity is the highest, and the linear capacity of the scenario that minimally violates the best case criterion. 
A number of interesting related questions remain open, among which a few are mentioned below. The achievable scheme of Theorem \ref{thm:1/2rate} uses scalar codes (the secret has only 1 symbol) while if block codes are used, the field size required can be reduced and the tradeoff between block-length and field size is an interesting problem.
As another example, while the best case scenarios are fully characterized, we know very little about the worst case scenarios, i.e., for which problem instances, the capacity is small and how small can it be? It is desirable to establish capacity approximations and exact capacity results for various classes of problem instances (e.g., in terms of the characteristic graphs). 
We have focused exclusively on the metric of capacity in this work, while other metrics are also interesting, e.g., the capacity region, the maximum randomness rate and the randomness constrained capacity. Extensions to include a larger number of users (beyond 2 users holding the secret) and more secrets (beyond a single secret) look fertile. %and in particular, the interplay with multiple secrets and finer access control structures introduces richer tradeoffs between various spaces involved in the problem and might necessitate new tools.
To sum up, this work represents an interesting initial step towards using signal overlap analysis and tools in information theory to understand the fundamental limits of multi-user primitives in cryptography, for which the potential remains promising while the topic is widely under-explored.

\let\url\nolinkurl
\bibliographystyle{IEEEtran}
\bibliography{Thesis}

\begin{thebibliography}{10}
\providecommand{\url}[1]{#1}
\csname url@samestyle\endcsname
\providecommand{\newblock}{\relax}
\providecommand{\bibinfo}[2]{#2}
\providecommand{\BIBentrySTDinterwordspacing}{\spaceskip=0pt\relax}
\providecommand{\BIBentryALTinterwordstretchfactor}{4}
\providecommand{\BIBentryALTinterwordspacing}{\spaceskip=\fontdimen2\font plus
\BIBentryALTinterwordstretchfactor\fontdimen3\font minus
  \fontdimen4\font\relax}
\providecommand{\BIBforeignlanguage}[2]{{%
\expandafter\ifx\csname l@#1\endcsname\relax
\typeout{** WARNING: IEEEtran.bst: No hyphenation pattern has been}%
\typeout{** loaded for the language `#1'. Using the pattern for}%
\typeout{** the default language instead.}%
\else
\language=\csname l@#1\endcsname
\fi
#2}}
\providecommand{\BIBdecl}{\relax}
\BIBdecl

\bibitem{shannon1949}
C.~E. Shannon, ``Communication theory of secrecy systems,'' \emph{Bell system
  technical journal}, vol.~28, no.~4, pp. 656--715, 1949.

\bibitem{Crypto_book}
J.~Katz and Y.~Lindell, \emph{Introduction to modern cryptography}.\hskip 1em
  plus 0.5em minus 0.4em\relax Chapman and Hall/CRC, 2014.

\bibitem{Devet_Goldberg}
C.~Devet and I.~Goldberg, ``The best of both worlds: Combining
  information-theoretic and computational pir for communication efficiency,''
  in \emph{International Symposium on Privacy Enhancing Technologies
  Symposium}.\hskip 1em plus 0.5em minus 0.4em\relax Springer, 2014, pp.
  63--82.

\bibitem{cramer_damgård_nielsen_2015}
R.~Cramer, I.~B. Damgård, and J.~B. Nielsen, \emph{Secure Multiparty
  Computation and Secret Sharing}.\hskip 1em plus 0.5em minus 0.4em\relax
  Cambridge University Press, 2015.

\bibitem{liang2009information}
Y.~Liang, H.~V. Poor, and S.~Shamai, ``Information theoretic security,''
  \emph{Foundations and Trends{\textregistered} in Communications and
  Information Theory}, vol.~5, no. 4--5, pp. 355--580, 2009.

\bibitem{bloch2011physical}
M.~Bloch and J.~Barros, \emph{Physical-layer security: from information theory
  to security engineering}.\hskip 1em plus 0.5em minus 0.4em\relax Cambridge
  University Press, 2011.

\bibitem{yener2015wireless}
A.~Yener and S.~Ulukus, ``Wireless physical-layer security: Lessons learned
  from information theory,'' \emph{Proceedings of the IEEE}, vol. 103, no.~10,
  pp. 1814--1825, 2015.

\bibitem{Sun_Jafar_PIR}
H.~Sun and S.~A. Jafar, ``{The Capacity of Private Information Retrieval},''
  \emph{IEEE Transactions on Information Theory}, vol.~63, no.~7, pp.
  4075--4088, 2017.

\bibitem{Banawan_Ulukus}
K.~Banawan and S.~Ulukus, ``{The Capacity of Private Information Retrieval from
  Coded Databases},'' \emph{IEEE Transactions on Information Theory}, vol.~64,
  no.~3, pp. 1945--1956, 2018.

\bibitem{Lee_Abbe}
E.~J. Lee and E.~Abbe, ``Two shannon-type problems on secure multi-party
  computations,'' in \emph{2014 52nd Annual Allerton Conference on
  Communication, Control, and Computing (Allerton)}.\hskip 1em plus 0.5em minus
  0.4em\relax IEEE, pp. 1287--1293.

\bibitem{Data_Prabhakaran_Prabhakaran}
D.~Data, V.~M. Prabhakaran, and M.~M. Prabhakaran, ``Communication and
  randomness lower bounds for secure computation,'' \emph{IEEE Transactions on
  Information Theory}, vol.~62, no.~7, pp. 3901--3929, 2016.

\bibitem{Zhou_Sun_Fu}
Y.~{Zhou}, H.~{Sun}, and S.~{Fu}, ``{On the Randomness Cost of Linear Secure
  Computation},'' in \emph{2019 53rd Annual Conference on Information Sciences
  and Systems (CISS)}, March 2019, pp. 1--6.

\bibitem{Zhao_Sun_SMP}
Y.~Zhao and H.~Sun, ``{Expand-and-Randomize: An Algebraic Approach to Secure
  Computation},'' \emph{arXiv preprint arXiv:2001.00539}, 2020.

\bibitem{yu2018lagrange}
Q.~Yu, S.~Li, N.~Raviv, S.~M.~M. Kalan, M.~Soltanolkotabi, and S.~Avestimehr,
  ``Lagrange coded computing: Optimal design for resiliency, security and
  privacy,'' \emph{arXiv preprint arXiv:1806.00939}, 2018.

\bibitem{chang2018capacity}
W.-T. Chang and R.~Tandon, ``{On the Capacity of Secure Distributed Matrix
  Multiplication},'' in \emph{2018 IEEE Global Communications Conference
  (GLOBECOM)}.\hskip 1em plus 0.5em minus 0.4em\relax IEEE, 2018, pp. 1--6.

\bibitem{Sun_Anonymous}
H.~Sun, ``The capacity of anonymous communications,'' \emph{IEEE Transactions
  on Information Theory}, vol.~65, no.~6, pp. 3871--3879, 2018.

\bibitem{Tahmasebi_Maddah}
B.~Tahmasebi and M.~A. Maddah-Ali, ``{Private Sequential Function
  Computation},'' \emph{arXiv preprint arXiv:1908.01204}, 2019.

\bibitem{Wang_Banawan_Ulukus}
Z.~Wang, K.~Banawan, and S.~Ulukus, ``{Private Set Intersection: A
  Multi-Message Symmetric Private Information Retrieval Perspective},''
  \emph{arXiv preprint arXiv:1912.13501}, 2020.

\bibitem{SymPIR}
Y.~Gertner, Y.~Ishai, E.~Kushilevitz, and T.~Malkin, ``Protecting data privacy
  in private information retrieval schemes,'' in \emph{Proceedings of the
  thirtieth annual ACM symposium on Theory of computing}.\hskip 1em plus 0.5em
  minus 0.4em\relax ACM, 1998, pp. 151--160.

\bibitem{Gay_Kerenidis_Wee}
R.~Gay, I.~Kerenidis, and H.~Wee, ``Communication complexity of conditional
  disclosure of secrets and attribute-based encryption,'' in \emph{Annual
  Cryptology Conference}.\hskip 1em plus 0.5em minus 0.4em\relax Springer,
  2015, pp. 485--502.

\bibitem{Applebaum_Arkis_Raykov_Vasudevan}
B.~Applebaum, B.~Arkis, P.~Raykov, and P.~N. Vasudevan, ``Conditional
  disclosure of secrets: Amplification, closure, amortization, lower-bounds,
  and separations,'' in \emph{Annual International Cryptology
  Conference}.\hskip 1em plus 0.5em minus 0.4em\relax Springer, 2017, pp.
  727--757.

\bibitem{Vaikuntanathan}
V.~Vaikuntanathan, ``Some open problems in information-theoretic
  cryptography,'' in \emph{37th IARCS Annual Conference on Foundations of
  Software Technology and Theoretical Computer Science (FSTTCS 2017)}.\hskip
  1em plus 0.5em minus 0.4em\relax Schloss Dagstuhl-Leibniz-Zentrum fuer
  Informatik, 2018.

\bibitem{Beimel_Survey}
A.~Beimel, ``Secret-sharing schemes: a survey,'' in \emph{International
  Conference on Coding and Cryptology}.\hskip 1em plus 0.5em minus 0.4em\relax
  Springer, 2011, pp. 11--46.

\bibitem{Laur_Lipmaa}
S.~Laur and H.~Lipmaa, ``A new protocol for conditional disclosure of secrets
  and its applications,'' in \emph{International Conference on Applied
  Cryptography and Network Security}.\hskip 1em plus 0.5em minus 0.4em\relax
  Springer, 2007, pp. 207--225.

\bibitem{Liu_Vaikuntanathan_Wee}
T.~Liu, V.~Vaikuntanathan, and H.~Wee, ``Conditional disclosure of secrets via
  non-linear reconstruction,'' in \emph{Annual International Cryptology
  Conference}.\hskip 1em plus 0.5em minus 0.4em\relax Springer, 2017, pp.
  758--790.

\bibitem{Applebaum_Vasudevan}
B.~Applebaum and P.~N. Vasudevan, ``Placing conditional disclosure of secrets
  in the communication complexity universe,'' in \emph{10th Innovations in
  Theoretical Computer Science Conference (ITCS 2019)}.\hskip 1em plus 0.5em
  minus 0.4em\relax Schloss Dagstuhl-Leibniz-Zentrum fuer Informatik, 2018.

\bibitem{Jafar_FnT}
\BIBentryALTinterwordspacing
S.~A. Jafar, ``{Interference Alignment - A New Look at Signal Dimensions in a
  Communication Network},'' \emph{Foundations and Trends in Communications and
  Information Theory}, vol.~7, no.~1, pp. 1--134, 2011. [Online]. Available:
  \url{http://dx.doi.org/10.1561/0100000047}
\BIBentrySTDinterwordspacing

\bibitem{Cadambe_Jafar_int}
V.~Cadambe and S.~Jafar, ``{Interference Alignment and the Degrees of Freedom
  of the $K$ user Interference Channel},'' \emph{IEEE Transactions on
  Information Theory}, vol.~54, no.~8, pp. 3425--3441, Aug. 2008.

\bibitem{MMK}
M.~{Maddah-Ali}, A.~Motahari, and A.~Khandani, ``Communication over {MIMO X}
  channels: Interference alignment, decomposition, and performance analysis,''
  in \emph{IEEE Trans. on Information Theory}, August 2008, pp. 3457--3470.

\bibitem{Wu_Dimakis}
Y.~Wu and A.~G. Dimakis, ``Reducing repair traffic for erasure coding-based
  storage via interference alignment,'' in \emph{Information Theory, 2009. ISIT
  2009. IEEE International Symposium on}.\hskip 1em plus 0.5em minus
  0.4em\relax IEEE, 2009, pp. 2276--2280.

\bibitem{Shah_IA}
N.~B. Shah, K.~Rashmi, P.~V. Kumar, and K.~Ramchandran, ``{Interference
  Alignment in Regenerating Codes for Distributed Storage: Necessity and Code
  Constructions},'' \emph{IEEE Transactions on Information Theory}, vol.~58,
  no.~4, pp. 2134--2158, 2012.

\bibitem{Cadambe_Jafar_Maleki_Ramchandran_Suh}
V.~Cadambe, S.~Jafar, H.~Maleki, K.~Ramchandran, and C.~Suh, ``Asymptotic
  interference alignment for optimal repair of mds codes in distributed data
  storage,'' \emph{IEEE Trans. on Information Theory}, vol.~59, no.~5, pp.
  2974--2987, May 2013.

\bibitem{meng2014precoding}
C.~Meng, A.~K. Das, A.~Ramakrishnan, S.~A. Jafar, A.~Markopoulou, and
  S.~Vishwanath, ``Precoding-based network alignment for three unicast
  sessions,'' \emph{IEEE Transactions on Information Theory}, vol.~61, no.~1,
  pp. 426--451, 2014.

\bibitem{Han_Wang_Shroff}
J.~Han, C.-C. Wang, and N.~B. Shroff, ``{Graph-theoretic Characterization of
  the Feasibility of the Precoding-based 3-unicast Interference Alignment
  Scheme},'' \emph{arXiv preprint arXiv:1305.0503}, 2013.

\bibitem{Maleki_Cadambe_Jafar}
H.~Maleki, V.~Cadambe, and S.~Jafar, ``{Index Coding -- An Interference
  Alignment Perspective},'' \emph{IEEE Transactions on Information Theory},
  vol.~60, no.~9, pp. 5402--5432, Sep. 2014.

\bibitem{Sun_Jafar_nonshannon}
H.~Sun and S.~A. Jafar, ``{Index Coding Capacity: How far can one go with only
  Shannon Inequalities?}'' \emph{IEEE Trans. on Inf. Theory}, vol.~61, no.~6,
  pp. 3041--3055, 2015.

\bibitem{Sun_Jafar_BIAPIR}
------, ``{Blind Interference Alignment for Private Information Retrieval},''
  in \emph{2016 IEEE International Symposium on Information Theory
  (ISIT)}.\hskip 1em plus 0.5em minus 0.4em\relax IEEE, 2016, pp. 560--564.

\bibitem{Jia_Sun_Jafar}
Z.~Jia, H.~Sun, and S.~A. Jafar, ``{Cross Subspace Alignment and the Asymptotic
  Capacity of X--Secure T--Private Information Retrieval},'' \emph{IEEE
  Transactions on Information Theory}, vol.~65, no.~9, pp. 5783--5798, 2019.

\bibitem{Schrijver}
A.~Schrijver, \emph{Combinatorial optimization: polyhedra and
  efficiency}.\hskip 1em plus 0.5em minus 0.4em\relax Springer, 2003, vol.~24.

\bibitem{Y_ITNC}
R.~W. Yeung, \emph{Information Theory and Network Coding}.\hskip 1em plus 0.5em
  minus 0.4em\relax Springer, 2008.

\bibitem{XITIP}
``{\it Xitip: Information Theoretic Inequalities Prover},'' Available:
  http://xitip.epfl.ch/, 2007.

\bibitem{TPH}
C.~Tian, J.~S. Plank, and B.~Hurst, ``An open-source toolbox for computer-aided
  investigation on the fundamental limits of information systems, version
  0.1,'' https://github.com/ct2641/CAI/releases/tag/0.1, October 2019.

\end{thebibliography}
\end{document}